*Opinion*
# Graphene in the fight against malaria

Amine El Moutaouakil [1,*], Mohamed Belmoubarik [2], and Weng Kung Peng [2,*]

[1] Electrical Engineering Department, College of Engineering, UAE University, P.O. Box 15551, Al Ain, UAE;
[2] International Iberian Nanotechnology Laboratory, 4715-330 Braga, Portugal;
[*] Correspondence: a.elmoutaouakil@uaeu.ac.ae (A.E.M.), weng.kung@inl.int (W.K.P.)

**Abstract:** Malaria infection is a major public health worldwide, with millions of new cases creating huge direct (an indirect) economic losses annually. Thus, the availability of preventive methods and diagnostic solutions for malaria is critical to save many lives, especially in poor countries. The emergence of graphene materials provided researchers with a promising path for a variety of fields including the medical field to fight epidemics and pandemics. In this contribution, we discuss the key-enabling graphene-based technologies in the fight against malaria, and how they are driving the innovation especially in the preventive and diagnostic phases of this endemic. We also discuss issues on biocompatibility of the graphene-based nanomaterials in human beings.

Malaria is a major public health concern which continues to claim the lives of more than half a million people each year. In the fight against malaria (e.g., control and elimination), two stages remain crucial, i.e.: prevention and diagnosis, in achieving the United Nations Sustainable Development Goals of ending malaria by 2030 [1,2]. There have been recent key-enabling technologies falling into a few different categories, i.e., targeting Plasmodium falciparum histidine-rich protein 2 (PfHRP-2), parasite lactate dehydrogenase (pLDH), aldolase, glutamate dehydrogenase (GDH), the biocrystal hemozoin, and biomolecules (e.g., DNA and proteins) [3–6]. In this perspective article, we discuss the emergence of graphene-based films and applications as key enabling technological innovations in both the preventive and diagnostic phases, that potentially allow the elimination and the control of malaria spread in global context [5,7–10].

## 1. Graphene nanomaterial: properties

Graphene [11,12] is a carbon-based 2D nanomaterial that is considered as a promising candidate for several, and somehow unrelated fields. First works on graphene highlighted its advantages over the widely used silicon material in today's electronic, but gradually shifted to emphasize it as a wonder material that can be used almost everywhere, from inks and flexible electronics to fabrics. Besides its cost-effectiveness and ease of synthesis, and thanks to its unique physical, mechanical, electrical, and optical features, graphene, as well as graphene-based devices, covered a variety of applications such as communication, electronic and medical applications to replace widely used nanomaterials (i.e. silicon and compound semiconductors) [13–27].

Graphene biosensing's features for viral infection have been thoroughly reported in the literature [19,28,29], and yet, it is believed that graphene can serve in both the prevention and the biosensing of parasitic infections such as malaria (Figure 1). Having a hyper-dimensional structure covering 1D, 2D and 3D, it has a wide surface area of 2630 m$^2$/g with an electrical conductivity exceeding 1000 S/m, and a thermal conductivity of 3000–5000 W/m.K. Its mechanical strength, Young's modulus of ~1.0 TPa, tunable bandgap and tensile strength not only satisfies the requirement of a biosensor, but also that of a barrier shield. Another advantage of graphene is the ability to synthesize it on different substrates and surfaces using various methods to yield different graphene-based materials, such as pristine graphene, graphene oxide (GO), reduced GO (rGO), and graphene quantum dots.

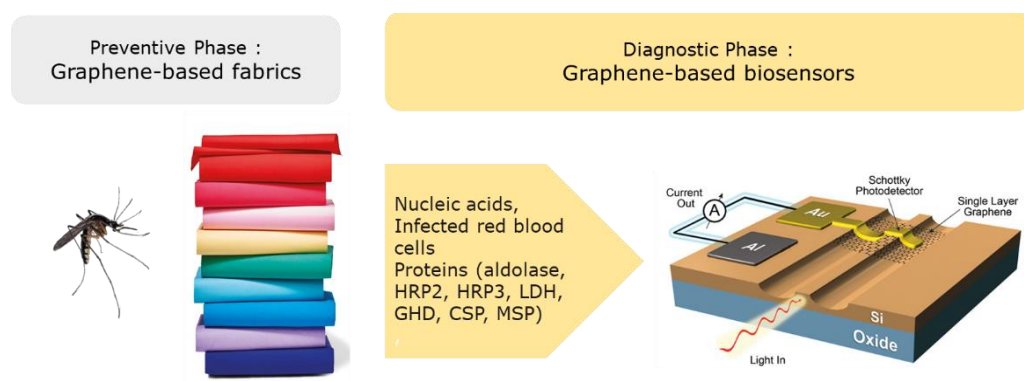

**Figure 1.** Fight against malaria - the role of graphene-based materials in the preventive and diagnostic phase. (a) Graphene-based films is used as barrier in fabrics which has the ability to interfere with the chemo-sensing. This is considered as a form of ´molecular barrier´ preventing mosquitoes from detecting skin-associated molecular attractants. (b) Graphene-based biosensors are designed to detect the presence of parasites (e.g., DNA) or the host biomolecule markers (e.g., proteins). Edited from [30].

## 2. Preventative phase:

### 2.1. Graphene-based nanomaterials in fabrics

Mosquitos are known to track humans from distances more than 50 meters thanks to their developed senses of thermal, visual and olfactory signals [31]. Humans can therefore be detected by their skin-associated molecular attractants, such as water vapor, $CO_2$, and skin microbiota metabolites. Approaches on how to hide these attractants are very important in the prevention against malaria.

Traditionally, the prevention against malaria is predominantly based on the long-lasting insecticidal nets (LLINs), considered one of the main vector-control strategies backed by WHO [2]. In high-risk areas, the population, especially women and children less than 5-year old, are recommended to sleep under LLINs to reduce the chances of mosquito bites. Other preventive strategies include chemoprevention -through the ingestion of drugs used against infections-, insecticide-treated nets (ITN), and indoor residual spraying (IRS). However, ITN and IRS have shown limited efficiency over time due to the resistance of parasites to the used chemicals [32]. The limited use of LLINs during rest/sleep time leaves the population vulnerable to mosquito bites during the rest of the day.

To face the robust detection multi-strategies and approaches of mosquitos, a robust camouflaging solution seems the only way to reduce the viral contamination. Graphene multilayer films on the other hand can serve as physical barrier to mosquito bites as well as a molecular barrier to conceal the skin-associated molecular attractants, thanks to their solid 2D structures, and their ability to withstand penetration forces of more than 50 μN in the case of 0.5-μm-thick GO films (Figure 2) [33]. Castilho et al. [33] have shown that using wearable technologies with embedded graphene-based films yields excellent results in the protection against mosquito bites as physical and molecular barriers at the same time. Dry GO and rGO films can interfere with the host-sensing system in mosquitos, especially A. aegypti, through the concealment of the skin attractants. In wet rGO films, where a mosquito attractant was introduced intentionally, 0.5-μm-thick graphene films showed sufficient mechanical puncture resistance against bite protection. Despite their lower biting frequency compared to cheesecloth, wet GO films however were easily penetrated by the mosquito bites due to their hygroscopic nature, making them easy to absorb water and swell, as well as easily destroyed during handling.

This research shows that graphene can play a big role in the prevention against malaria in several dimensions; The integration of graphene-based films in the wearable technologies can decrease the frequency and the chances of mosquito bites during all-day outing and activities. Combined with the cheap-cost of graphene material, its non-toxicity, flexibility and ease of synthesis, this provides an additional, if not an alternative, layer of protection to the widely-used LLINs.

Other than clothing and LLINs-like nets with graphene-based materials, the use of graphene-based coating on the walls of housings in mosquitos-populated area can provide a large-scale protection to the inhabitants against mosquito bites, as this will serve as a chemical barrier and as a camouflage to the $CO_2$ plume produced by the residents, and thus deprive mosquitos of one of their long-distance sensing weapons. This alone will reduce the density of mosquitos inside the housing premises, and therefore provide a guaranteed reduction in the spread of malaria.

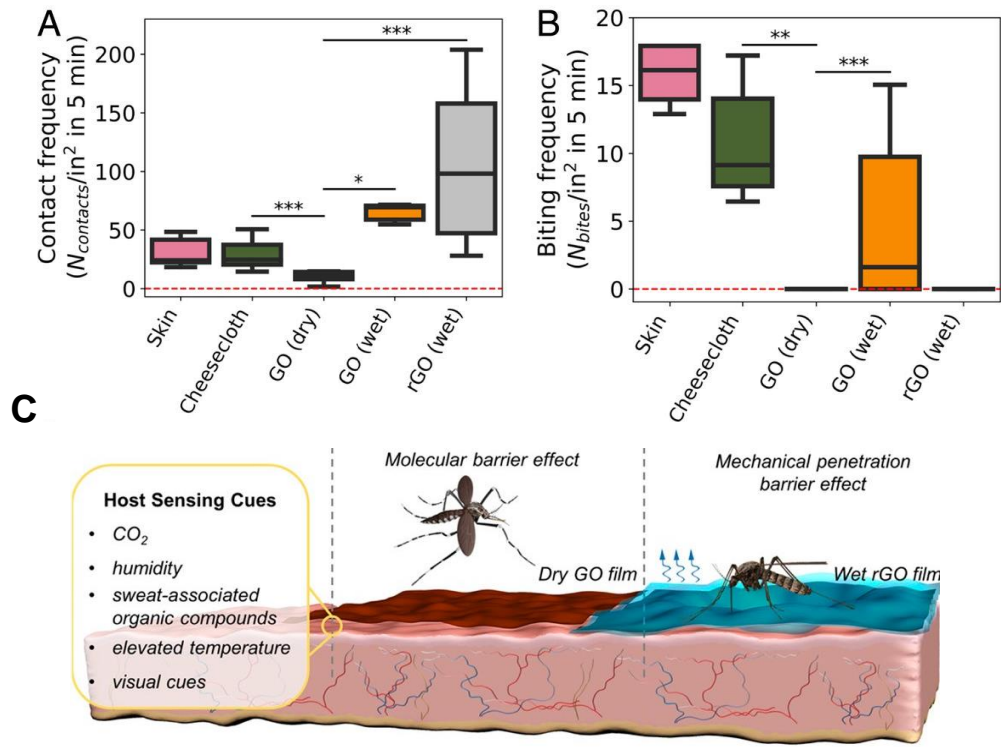

**Figure 2.** Mechanism of mosquito bite inhibition. (A) The landings plus walk-ons of mosquitos on dry and wet GO-rGO films and controls (cheesecloth and skin). (B) The mosquito residence times after initial contact on dry and wet GO-rGO films and controls. (C) Representation of mosquito bite inhibition mechanisms on dry GO and wet rGO films. Adapted from [33].

*2.2. Graphene-based nanomaterials in bloodstream*

Besides the wearable technologies, the use of graphene-based materials directly in human bloodstream showed barrier-like behavior similar to the ones highlighted in Section 2.1 [7]; The hydrophilic GO was found to be more suitable for bio-applications, as opposed to the hydrophobic pristine graphene, thanks to its lower toxicity and higher biocompatibility. In-vitro investigations [7] on the interactions between different concentrations of GO and Plasmodium falciparum (P. falciparum), a specie of malaria parasites, have shown that GO nanosheets can serve as an efficient physical obstruction and barrier between red blood cells (RBCs) and the parasite. The merozoites, resulting from the parasite development within the human bloodstream, didn't have access to healthy RBCs due to the limited access resulting from their adhesion to the GO nanosheets. The growth of the parasites from rings to trophozoites was also delayed and their maturity was delayed when GO was introduced in infected RBCs, showing the antimalarial behavior of GO. Other than GOs, graphene quantum dots (GQDs) showed excellent chemical and physical protperties for biomedical application, and their toxicity against P. faciparum, both CQ-resistant and CQ-sensitive strains, P. berghei and young instars of malaria mosquitoes were demonstrated [34]. The GQDs showed also toxicity against MCF-7 breast cancer cell [34].

This result clearly shows the potential of GO and other graphene-based nanomaterials in preventing the prolificaton of malaria parasites in the bloodstream, as well as inhibiting their development into advanced stages, which opens the way in the development of new nanomaterial-based approaches for fighting malaria.

*2.3. Bio-impact of graphene-based nanomaterials on humans*

While the toxicity and bio-impact of graphene-based materials were discussed in several literature [7,33,34], they are still one of the pending issues in the integration of graphene-based nanomaterials in the prevention solution against malaria. The bio-assessment of graphene-based materials has been one of the research topics gaining more popularity due to the ever-increasing interest in graphene and its derivatives. Yet, the potential impact of these nanomaterials on humans and on the environment is still lacking data, especially in vivo, due to the shortage of robust and validated approaches for toxicology testing [14], which will make the prediction of toxicity based solely on the material properties of the materials.

On the bio-impact of graphene-based materials, there are only limited toxicological data at the skin level, with only one in vivo study, while the majority are in vitro studies on skin fibroblasts and/or keratinocytes depending on different derivatives. The graphene, for instance, was found to induce higher cytotoxicity on skin fibroblasts than GO due to the material aggregation [35], while all graphene-based materials induced significant cytotoxicity on human keratinocytes with different potencies based on the oxidative state of the materials [36]. The only available in vivo investigation showed that the GO injected into the dermis of the growing feather sites of chickens, initiated an immune response after dermal injection [37]; From 3 days onwards, clear aggregates of the injected graphene-based material could be observed, and appeared to be inside cells or part of build-ups of cells inside and around the nanomaterial as shown in Figure 3.

Nevertheless, the integration of graphene-based materials in wearable technologies is less invasive compared to the mentioned studies, due to the absence of direct contact with body cells, but we can consider skin irritation the most logical outcome after cutaneous exposure [14]. This can be solved through the use of graphene derivatives as inner layers in fabrics rather than on the surface, to suppress any dermal contact.

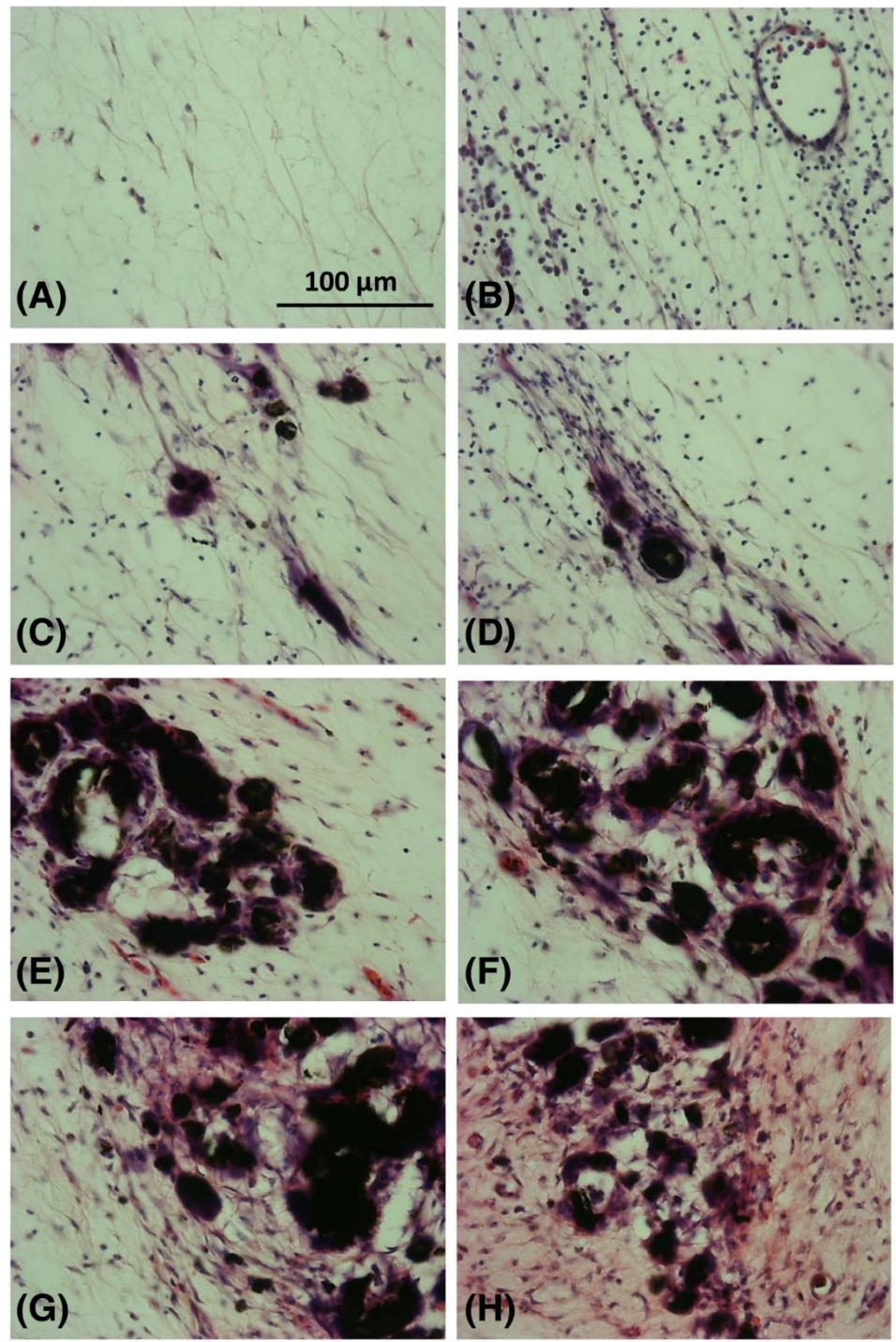

**Figure 3.** Optical microscopy images of eosin- and hematoxylin-marked fixed tissue sections prepared from growing feathers injected with either graphene-based nanomaterials (seen in black) or phosphate-buffered saline (PBS) vehicle. (A) before the injection; (B) at 1 day-post-injection, (C–F) at 3 days post-injection, and (G & H) at 5 days post-injection. Adapted from [37]

## 3. Diagnostic phase: Graphene-based biosensor

In the absence of appropriate preventive measures, early-stage diagnostic of malaria is considered a key element in the fight against malaria, as it can prevent the transmission, reduce the draining of the available resources, minimize the unnecessary use of drugs, avoid the spread of life-threatening side-effects, as well as the spread of drug resistance among the population [2]. Diagnosis has mainly been used through the use of antigen-based rapid diagnostic tests (RDTs) that remain expensive and not as sensitive as the thick blood films, which require advanced equipment.

With an annual death rate of more than one million in Africa alone, there is a vital need for efficient, easy-to-deploy, cost-effective and simple-to-operate detection methods to help in the malaria fight. Graphene-based biosensors are promising candidates for the electrochemical detection and new generation point-of-care testing thanks to the graphene electrical and optical properties [8,9,38].

The advanced electrical and optical properties available in the graphene has been exploited in several areas [11,12,39,40]. One of the area that are attractive for graphene biosensing is the disposable screen-printed electrodes (SPEs), where the use of the less performant multi-wall carbon nanotubes (MWCNTs) showed excellent results; a 96% sensitivity and a 94% specificity [41]. The integration of graphene-based materials instead of MWCNTs will provide a better sensitivity and higher signal-to-noise ratios than the reported sensor. Besides, the flexible nature of the graphene opens the way for the graphene-based electrodes to be printed on other graphene-based flexible substrates and be integrated in printed-circuit boards (PCB) to offer commercially viable options for malaria biosensors in bedside applications.

In addition, graphene-based sensors can integrate both the RDT and thick blood films thanks to their ability of monitoring the electronic transfer reactions of haemoglobin concentration, and thus check for the existance of malaria-caused anaemia [8,9]. The iron ion inside the haemoglobin can be in $Fe^{II}$ or $Fe^{III}$ oxidation states, the latter, which is caused by malaria, hinders methemoglobin from binding to oxygen, resulting in fatality in the absence of internal reduction mechanisms within the the RBCs. In fact, Toh. R.J. et al has shown that it is possible to make direct in vivo of electrochemical detection of haemoglobin in RBCs on glassy carbon electrodes using Naftion (Figure 4). This study confirms that the multifaceted biological conditions of a human RBC showed no interference with the sensing of haemoglobin. This clearly show the possibility of studying the electrochemical behavior of haemoglobin directly from human blood [8].

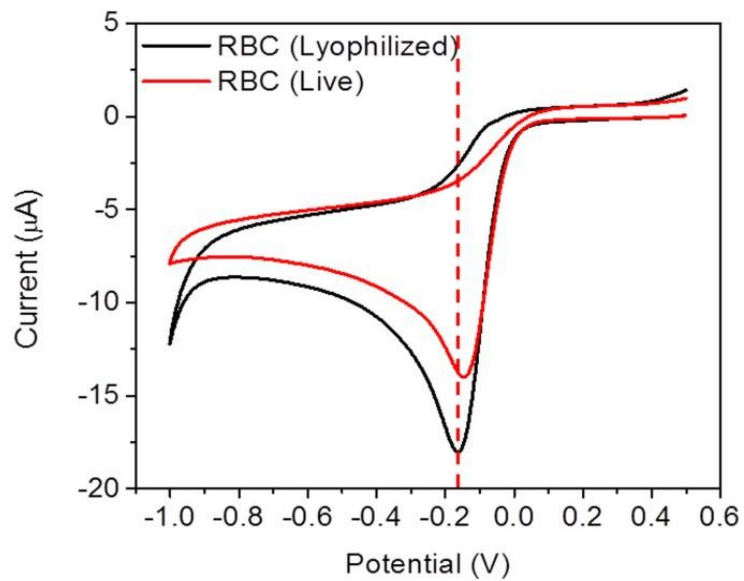

**Figure 4.** Study of the reduction of live RBCs (obtained directly from a real blood sample) compared to lyophilized RBCs (from Sigma Aldrich) using cyclic voltammetry (CV). The similar reduction potential points and shape of the CV indicates the possibility of detecting haemoglobin acitivity without much signal nose. Adapted from [8].

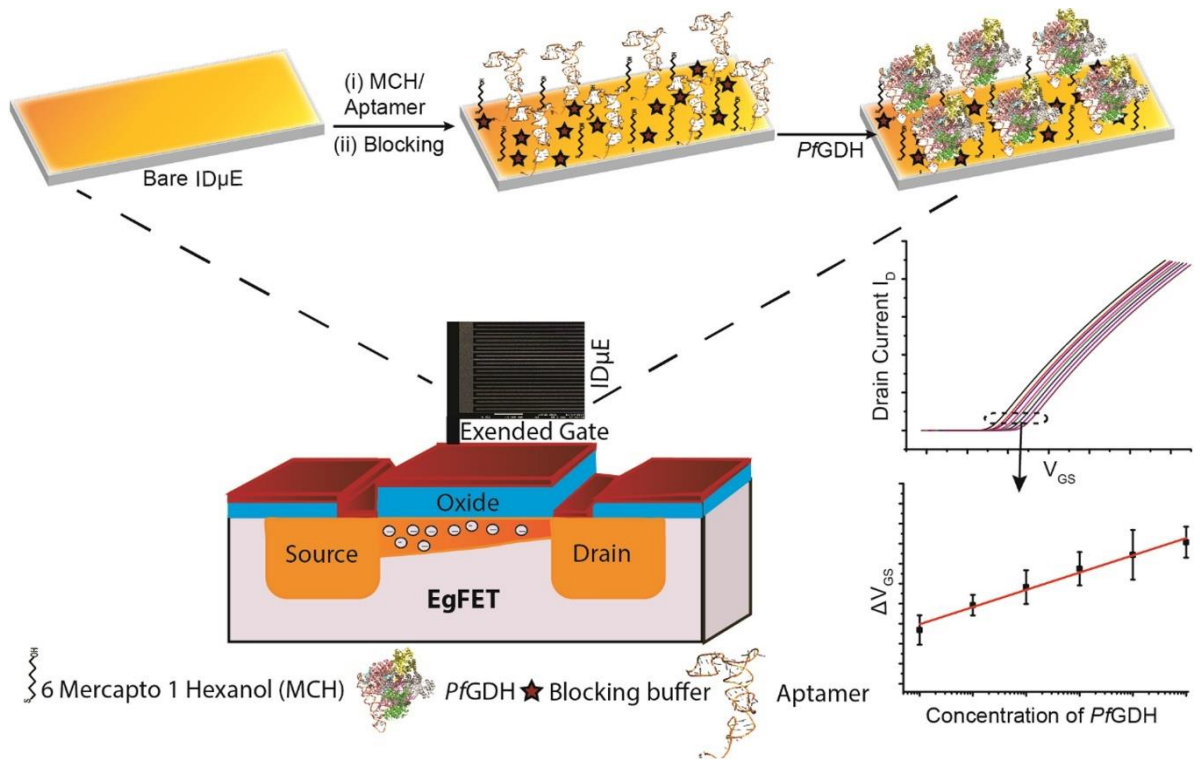

**Figure 5.** Process of fabrication for aptaFET for detection of the malaria biomarker PfGDH in blood; the detection is made using an extended-gate FET (EgFET) equipped with a selective aptamer as the biorecognition element against the PfGDH biomarker. The gold-thiol chemistry on inter-digitated gold microelectrode (IDµE) connected to the gate of transistor was used to immobilize the aptamer. Adapted from [42].

Moreover, graphene derivatives was found to strengthen and enhance the performance of existing solutions such as the glassy carbon; Even if the latter is preferred over other metals due to its high hydrogen overpotential, graphene-modified glassy carbon electrodes provide better selectivity in isomers, that has the same electro-active groups with overlapping redox signatures [43]; each atom in a graphene or graphene-based sheet is a surface atom, increasing the sensitivity of the electron transport and molecular interaction with adsorbed molecules. Therefore, electrodes coated with graphene or its derivatives can provide higher selectivity and higher sensitivity in the direct detection of haemoglobin from RBCs and blood samples, which will allow for quantitative analysis of RBCs in blood samples for clinical diagnosis of malaria.

The advances in field effect transistors (FET) and their integration in the biomedical field as high-sensitivity biosensors [44–46] is opening also the path for graphene and its derivatives to be incorporated as channel material instead of the technologically-limited silicon materials; in fact, aptamer-based field effect transistor (aptaFET) with silicon materials showed a higher sensitivity and selectivity to detect Plasmodium falciparum glutamate dehydrogenase (PfGDH) in blood, by measuring the changes in the gate potential in the FET as shown in Figure 5 [42]. Repalcing silicon material in post moore's law era with a graphene-based semiconductors will enhance further these devices and provide more biosensing capabilities [19].

## 4. Future challenges and opportunities

Graphene and its derivative nanomaterials have attracted a huge interest in several fields including the biomedical area. Applications in the medical domain ranges from regenerative medicine to gene and drug delivery and cancer therapy [47]. The antimicrobial feature of the graphene-based materials, however, needs more research and investigation due to the reported aggregation and mode of exposure effect on graphene and related materials' cytotoxicity [47]. The oxygen content, the dosage, surface functional groups, structural defects are all parameters that are reported to affect the toxicity of this material. The bio-impact of graphene-based materials is also a research topic that needs more data and relevant approaches, especially through in vivo investigation.

In the fight against malaria, graphene-based devices can cover both the preventive and the diagnosis phases; but one area is becoming more promising; the DNA detection using graphene-based materials [48]. DNA detection is considered an attractive application in disease diagnosis, including malaria, but there are two challenges facing this field; the speed and the accuracy of the detection. Mainly the simple and less-expensive label-free DNA detection using graphene derivatives is attractive thanks to the graphene properties, which enable it to be integrated in optical, electronic, and electrochemical biosensors. For instance, the integration of graphene in electronic biosensors can be made through the use of the DNA in the gating area, changing the transfer characteristics as a result of the n-doping phenomenon (Figure 6) and showing higher sensitivity compared to available biosensors [49]. The use of recognition elements such as the DNA-mimicking PNA opens the way toward ultra-sensitive DNA detection (Figure 6 (d)).

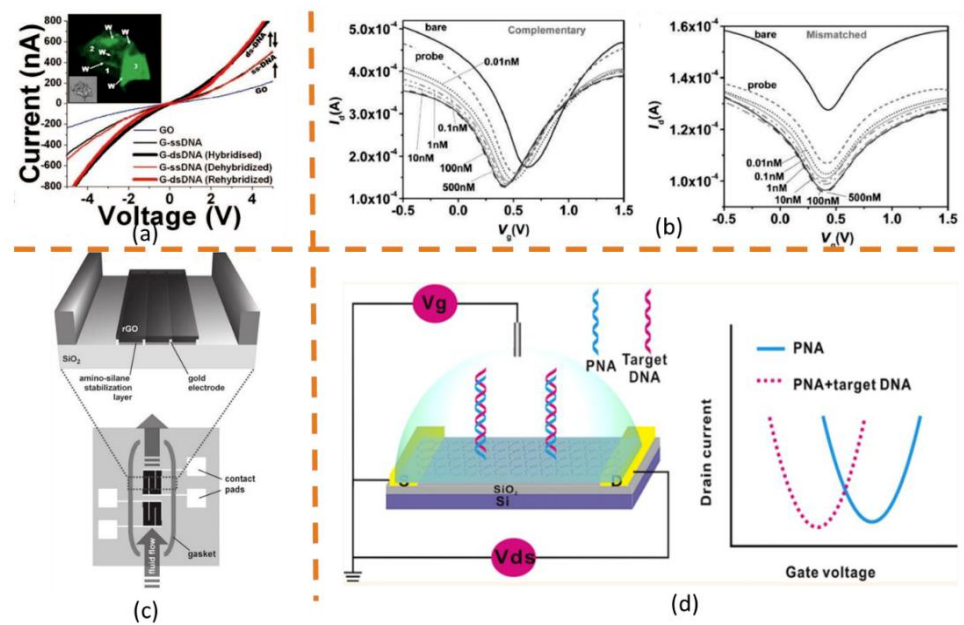

**Figure 6.** (a) Graphene-based DNA transistor and the effect of DNA hybridization and dehybridization; (b) Transfer characteristics before and after adding DNA with the concentration ranging from 0.01 to 500 nM; (c) Schematic representation of the sensor with a reference part; (d) Schematic of the rGO FET biosensor for DNA detection based on PNA-DNA hybridization. Adapted from [48].

The challenges that are facing the DNA detection are related to the DNA translocation and the noise level encountered during the detection. The availability of different graphene derivatives with a wide range of properties, and with better control over modulations in the tunneling current within the graphene can solve these issues and provide high sensitivity and selectivity in DNA biosensors.

The use of graphene-related nanomaterials is launching a new era in the fight against malaria; preventive measures can be enhanced through the use of these materials in the new wearable technologies, which will provide more efficient and cheaper solutions. Diagnosis of the malaria will also be enriched with compact and easy-to-deploy solutions such as electrochemical detection of haemoglobin, as well as DNA detection device based on graphene and its derivates. Moreover, It may be possible to provide hybrid solution using graphene device for direct parasite detection in the DNA and the oxidation states of the haemoglobin. Of which, high sensitivity and high specificity of the test are desirable in the field settings [50,51].

## 5. Conclusions

We have discussed extensively the recent advances made in graphene-based technology in the fight against malaria, and discussed issues on graphene compatibility on human beings. These set of nanomaterials holds both preventive and diagnostic features that can be applied to help in the control and reduction of malaria. We hope that this article is able to spur interest into studies of graphene-based materials in multiple aspects of the fight against malaria.

**Declarations**

**Ethics approval and consent to participate:** Not applicable.

**Consent for publication:** Not applicable.

**Availability of data and materials:** Not applicable.

**Competing interests:** The authors declare no conflict of interest.

**Funding:** This work was supported by funds from the UPAR project No. 31N393. W.K. Peng would like to thanks to INL Start Up Grant S4000040 and INL Seed Grant 2018.

**Authors' contributions:** AEM and WKP designed the outline of the manuscript and all authors analyzed the available data and made the literature research. All authors contributed to the opinion development in this manuscript and AEM was a major contributor in writing the manuscript. All authors read and approved the final manuscript.

**Acknowledgements:** Not applicable.

11 of 1224. Moutaouakil AE, Suemitsu T, Otsuji T, Coquillat D, Knap W. Room temperature terahertz detection in high-electron-mobility transistor structure using InAlAs/InGaAs/InP material systems. 35th International Conference on Infrared, Millimeter, and Terahertz Waves [Internet]. Rome, Italy: IEEE; 2010 [cited 2020 Jun 7]. p. 1–2. Available from: http://ieeexplore.ieee.org/document/5612598/

25. Moutaouakil AE, Komori T, Horiike K, Suemitsu T, Otsuji T. Room Temperature Intense Terahertz Emission from a Dual Grating Gate Plasmon-Resonant Emitter Using InAlAs/InGaAs/InP Material Systems. IEICE Transactions on Electronics. 2010;E93.C:1286–9.

26. El Moutaouakil A, Suemitsu T, Otsuji T, Videlier H, Boubanga-Tombet S-A, Coquillat D, et al. Device loading effect on nonresonant detection of terahertz radiation in dual grating gate plasmon-resonant structure using InGaP/InGaAs/GaAs material systems. Phys Status Solidi (c). 2011;8:346–8.

27. Moutaouakil AE, Suemitsu T, Otsuji T, Coquillat D, Knap W. Nonresonant Detection of Terahertz Radiation in High-Electron-Mobility Transistor Structure Using InAlAs/InGaAs/InP Material Systems at Room Temperature. j nanosci nanotechnol. 2012;12:6737–40.

28. Bellan LM, Wu D, Langer RS. Current trends in nanobiosensor technology. Wiley Interdiscip Rev Nanomed Nanobiotechnol. 2011;3:229–46.

29. Suvarnaphaet P, Pechprasarn S. Graphene-Based Materials for Biosensors: A Review. Sensors. Multidisciplinary Digital Publishing Institute; 2017;17:2161.

30. Goykhman I, Sassi U, Desiatov B, Mazurski N, Milana S, de Fazio D, et al. On-Chip Integrated, Silicon–Graphene Plasmonic Schottky Photodetector with High Responsivity and Avalanche Photogain. Nano Lett. American Chemical Society; 2016;16:3005–13.

31. van Breugel F, Riffell J, Fairhall A, Dickinson MH. Mosquitoes Use Vision to Associate Odor Plumes with Thermal Targets. Current Biology. Elsevier; 2015;25:2123–9.

32. N'Guessan R, Corbel V, Akogbéto M, Rowland M. Reduced Efficacy of Insecticide-treated Nets and Indoor Residual Spraying for Malaria Control in Pyrethroid Resistance Area, Benin. Emerg Infect Dis. 2007;13:199–206.

33. Castilho CJ, Li D, Liu M, Liu Y, Gao H, Hurt RH. Mosquito bite prevention through graphene barrier layers. Proc Natl Acad Sci USA. 2019;116:18304–9.

34. Murugan K, Nataraj D, Jaganathan A, Dinesh D, Jayashanthini S, Samidoss CM, et al. Nanofabrication of Graphene Quantum Dots with High Toxicity Against Malaria Mosquitoes, Plasmodium falciparum and MCF-7 Cancer Cells: Impact on Predation of Non-target Tadpoles, Odonate Nymphs and Mosquito Fishes. J Clust Sci. 2017;28:393–411.

35. Liao K-H, Lin Y-S, Macosko CW, Haynes CL. Cytotoxicity of Graphene Oxide and Graphene in Human Erythrocytes and Skin Fibroblasts. ACS Appl Mater Interfaces. American Chemical Society; 2011;3:2607–15.

36. Pelin M, Fusco L, León V, Martín C, Criado A, Sosa S, et al. Differential cytotoxic effects of graphene and graphene oxide on skin keratinocytes. Scientific Reports. Nature Publishing Group; 2017;7:40572.

37. Erf GF, Falcon DM, Sullivan KS, Bourdo SE. T lymphocytes dominate local leukocyte infiltration in response to intradermal injection of functionalized graphene-based nanomaterial. Journal of Applied Toxicology. 2017;37:1317–24.